# YBCO-buffered NdBCO film with higher thermal stability in seeding REBCO Growth

H. H. Xu[1], Y. Y. Chen[1], L. Cheng[1], S. B. Yan[1], D. J. Yu[1], X. Yao[1, 2, *]

[1]Key Laboratory of Artificial Structures & Quantum Control (Ministry of Education)

 Department of Physics

[2]State Key Laboratory for Metal Matrix Composites,

Shanghai Jiao Tong University, 800 Dongchuan Road, Shanghai 200240, China

[*]Corresponding author

E-mail address: xyao@sjtu.edu.cn (X. Yao).

## ABSTRACT

In this work, we report a strengthened superheating effect caused by a buffering $YBa_2Cu_3O_y$ (Y123 or YBCO) layer in the $Nd_{1+x}Ba_{2-x}Cu_3O_{7-y}$ (Nd123 or NdBCO) thin film with MgO substrate (i.e., NdBCO/YBCO/MgO thin film). In the cold-seeding melt-textured (MT) growth, the NdBCO/YBCO/MgO film presented an even higher superheating level, about 20 °C higher than that of non-buffered NdBCO film (i.e., NdBCO/MgO film). Using this NdBCO/YBCO/MgO film as seeds and undergoing a maximum processing temperature ($T_{max}$) up to 1120 °C, we succeeded in growing various $RE_{1+x}Ba_{2-x}Cu_3O_{7-y}$ (REBCO, RE=rare elements) bulk superconductors, including $Gd_{1+x}Ba_{2-x}Cu_3O_{7-y}$ (GdBCO), $Sm_{1+x}Ba_{2-x}Cu_3O_{7-y}$ (SmBCO) and NdBCO that have high peritectic temperatures ($T_p$). The pole figure (X-Ray Φ-scan) measurement reveals that the NdBCO/YBCO/MgO film has better in-plane alignment than the NdBCO/MgO film, indicating that the induced intermediate layer improves the crystallinity of the NdBCO film, which could be the main origin of the enhanced thermal stability. In short, possessing higher thermal stability and enduring a higher $T_{max}$ in the MT process, the NdBCO/YBCO/MgO film is beneficial to the growth of bulk superconductors in two aspects: (1) broad application for high-$T_p$ REBCO materials; (2) effective suppression against heterogeneous nucleation, which is of great assistance in growing large and high-performance REBCO crystals.

I. Introduction

Nowadays, top-seeded melt-texture growth (TSMTG) is commonly used to obtain large single domain REBCO bulks. In this method, the seed crystal or film is placed on the top of the precursor to induce the epitaxial nucleation. The seed is required to: (1) have a similar crystal structure to REBCO; (2) remain stable during the whole TSMTG process. SmBCO and NdBCO crystals have been commonly used as seeds. However, since their peritectic temperatures are not high enough, the members they could seed in the REBCO family are limited.[1] The Mg-doping to NdBCO fascinatingly breaks this limitation, raising its $T_p$ by 20 °C over the $T_p$ of pure NdBCO.[2]

In our previous work, superheating phenomenon was demonstrated in REBCO thin films which were deposited on the MgO substrate. Firstly, a YBCO thin film, possessing a $T_p$ of 1010 °C, was found to be able to endure a high processing temperature (1057 °C) for inducing an epitaxial growth of an NdBCO ($T_p$=1085 °C) thick film in liquid phase epitaxial growth.[3] In the second example, where the superheating nature appears more evidently, a YBCO bulk was induced by using its homoseed, a YBCO/MgO film, after holding at a temperature of 30 °C above its $T_p$ for 1.5 hours.[4] Furthermore, the superheating phenomenon was clearly identified to exist in all commonly-used YBCO-film/substrate structure (where substrate is MgO, LaAlO$_3$ or SrTiO$_3$).[5] All these facts suggest a universal superheating mode, which relates to a distinctive construction feature absolutely, i.e., a free surface with low surface energy and a semi-coherent interface between film and substrate. So far, most kinds of REBCO bulks have been successfully seeded by NdBCO/MgO thin films, which become potentially popular because NdBCO has the highest $T_p$ among the commonly applied REBCO materials.[6] However, being in the structure of thin film, NdBCO has a superheating upper limit at about 1098 °C in MT growth, only 14 °C higher than its $T_p$, which is lower than that of a Y123 thin film.[7] Compared with the YBCO system, due to a lower liquidus slop, there is a higher Nd supersaturation for the RE123 growth in the same undercooling. Similarly, there is a higher Nd supersaturation for the RE$_2$BaCuO$_5$ (RE211) growth in the same superheating, which is equivalent to a higher growth rate of the Nd$_4$Ba$_2$Cu$_2$O$_5$ (Nd422, with a cation compositional ratio of RE211) and higher dissolution rate of the Nd123.[7] In addition, the manufacturer Theva GmbH reported that the growth window is rather small to obtain high-quality NdBCO/MgO films.[8] All these two demerits forbid the NdBCO/MgO film to seed the growth of high-$T_p$ REBCO bulks, e.g., RE=Sm or Nd. Recently, by growing a YBCO buffer between the NdBCO film and the MgO substrate, namely, an NdBCO/YBCO/MgO film, the manufacturer applauded for its effect on broadening the growth window of high-quality NdBCO films.[8]

With this special NdBCO-film/YBCO-buffer-layer/MgO-substrate structure, better crystallinity and higher thermal stability of the NdBCO film can be expected. In this work, using NdBCO/YBCO/MgO thin films as seeds in the MT process, a temperature as high as 1120 °C was tolerated, which was remarkably 35 °C higher than the $T_p$ of NdBCO. Seeded by this NdBCO/YBCO/MgO film, in particular, a fully-grown NdBCO bulk without Ag addition was successfully obtained.

## II. EXPERIMENTS

The NdBCO/YBCO/MgO thin film, with a 600-nm NdBCO film and a 25-nm YBCO buffer layer, was provided by the company Theva GmbH. The in-plane alignment for both NdBCO/YBCO/MgO and NdBCO/MgO thin films (also made by Theva GmbH) were tested by pole figure analysis using Cu Kα radiation, indicating the orientation relationship between the film and the MgO substrate. The Nd123 and Nd422 precursor powder were prepared by grinding and calcining a stoichiometric mixture of $Nd_2O_3$, $BaCO_3$ and CuO raw powders repeatedly. All the synthesized powders were characterized by the X-ray diffraction (XRD) method. Sample with the nominal compositions of Nd123 + 20 mol%Nd422 + 1 wt%$CeO_2$ was prepared and pressed uniaxially into pellet of 20 mm in diameter. The growth of NdBCO sample was performed in the air using a typical TSMTG method. The NdBCO/YBCO/MgO thin film was placed on the top of the pellet as a seed, to control the growth of a single grain in the (001) orientation. The heating profile was as follows: the sample was heated from room temperature to 950 °C in 5 hours, held at 950 °C for 4 hours, and then heated to 1117 °C in 2 hours. After held at 1117 °C for 1.5 hours, the sample was cooled rapidly to 1083 °C in 20 minutes, and then slowly cooled down at a rate of 0.25 °C h$^{-1}$ before quenching.

## III. RESULTS AND DISCUSSION

By cold-seeding an NdBCO/YBCO/MgO thin film to induce a homo-epitaxy, we report a success on the full growth of a single-grain NdBCO bulk without the Ag addition. Remarkably, this NdBCO/YBCO/MgO film seed has endured the $T_{max}$ of 1117 °C for 1.5 hours, which is about 20 °C higher than the $T_{max}$ (~1100 °C) that NdBCO/MgO thin film could endure.[7, 12, 13] It is evident that the NdBCO/YBCO/MgO thin film possesses a higher superheating upper limit. More importantly, using a higher $T_{max}$ in the MT process is of great assistance in growing large-size and high-performance REBCO bulks.[9] Figure 1 shows the top view of the as-grown NdBCO bulk with a size of 16 mm in diameter, presenting four-fold growth facet lines clearly.

Figure 2 shows the pole figure analysis for the two kinds of NdBCO films with and without a YBCO buffer layer. A noticeable improvement of crystallinity in the YBCO-buffered NdBCO film is exhibited.

The pole figure measurement result from the NdBCO/MgO thin film (Figure 2a), presents a poor in-plane alignment. There are small unexpected peaks (satellite peaks) at ±17.5° around the strong peaks, indicating that the <100> direction of the NdBCO thin film orients to the <100> direction of the MgO substrate in a 0° and (0±17.5)° relationship. In comparison, the result from the NdBCO/YBCO/MgO thin film (Figure 2b), reveals a good four-fold symmetry, indicating that there is only one in-plane orientation in the NdBCO/YBCO/MgO thin film, i.e., the <100> direction of the NdBCO/YBCO thin film orients well in a 0° relationship to the <100> direction of the MgO substrate. Convincingly, it is the YBCO intermediate layer that plays a role in improving the in-plane alignment of the NdBCO film. Similar to the sandwich structure of REBCO/buffer/substrate in REBCO coated conductors, the YBCO buffer layer in NdBCO/YBCO/MgO thin film performs functions in: (1) compensating the lattice misfit between NdBCO and MgO; (2) alleviating the problems related to the interface from the substrate.[10]

As we know, the melting always initiates at defect sites (e.g., free surfaces and internal grain boundaries) because of their high free energy, defects of grain boundaries in one film play an important role in the thermal stability. Due to the excellent in-plane alignment, the grain boundaries in the NdBCO/YBCO/MgO thin film are much fewer than those in the NdBCO/MgO thin film, which predominantly takes the responsibility for the enhanced superheating capability in the NdBCO/YBCO/MgO thin film.

Considering the similar crystal structure in the whole REBCO family, it is reasonable to assume that almost all kinds of REBCO bulks could be seeded by NdBCO/YBCO/MgO thin films. Table 1 lists the various kinds of REBCO bulks effectively seeded by the NdBCO/YBCO/MgO thin film and the $T_{max}$ of each process we succeeded in. Figure 3, as one more distinctive result, shows the top view of the GdBCO single grain induced by the YBCO-buffered NdBCO film seed undergoing a $T_{max}$ of 1120 °C, which is a $T_{max}$-tolerating record in the homo-seeding MT growth of REBCO bulk as far as we know.

Figure 4 shows the types of film seeds that have been practically used and the $T_{max}$ they have been reported in the MT processes. It can be seen that the intrinsic $T_p$ of REBCO significantly affects the $T_{max}$ a film could tolerate. The $T_{max}$ for YBCO, SmBCO and NdBCO film are 1045 °C [4], 1090 °C [11] and 1098~1100 °C [7, 12, 13] (the difference in the upper-limit temperature is due to different furnace conditions or temperature criterion) respectively, which are in a rising trend corresponding to the increasing $T_p$ for YBCO, SmBCO and NdBCO. What is more, the film structure affects the $T_{max}$ significantly, as the $T_{max}$ for NdBCO/YBCO/MgO film is about 20 °C higher than that for NdBCO/MgO film. In short, this unique NdBCO-film/YBCO-buffer-layer/MgO-substrate structure significantly improves the superheating effect, which is a very interesting phenomenon of great practical applications.

There are 3 kinds of surface-related defects that are relative to the thermal stability of one film: (1) hetero-structure interface, like MgO/NdBCO interface; (2) inner grain boundaries in the film; (3) exposed surface with hanged bonding. YBCO buffer layer inducing better in-plane alignment reduces inner grain boundaries in the NdBCO film. But there is some negative effect, too. Because of the great difference in the intrinsic $T_p$ between YBCO and NdBCO, YBCO buffer decomposes earlier than NdBCO. In the MT heating process, the YBCO buffer decomposes into the Y211 solid and the Ba-Cu-O (BCO) liquid, forming a BCO-liquid/NdBCO hetero-structure interface, which has a higher surface energy. However, the small Nd123 grains in the film might become coarsened during the long time heating, which also reduces the grain boundaries and slows down the nucleation rate according to Griffith et al. .[14] Besides, the over-saturation of rare earth element in the BCO liquid could suppress the decomposition of Nd123 grain, making the film seed stable in the MT process.[4] In other words, better in-plane alignment induced by YBCO buffer layer plays a dominant role in the enhanced thermal stability of the NdBCO/YBCO/MgO film. How these three factors affect thermal stability particularly and how their specific mechanisms work are still under investigation and will be discussed in another paper.

## IV. CONCLUSION

Using the NdBCO/YBCO/MgO thin film as a seed and taking a $T_{max}$ of 1117 °C in the MT process, an NdBCO bulk superconductor with a full growth was obtained. According to the X-ray pole figure results, NdBCO/YBCO/MgO thin film has better crystallinity than NdBCO/MgO thin film, which leads to the fact that the NdBCO/YBCO/MgO thin film possesses a stronger superheating effect than the NdBCO/MgO thin film. So a higher $T_{max}$ could be used during the MT growth seeded by the NdBCO/YBCO/MgO thin film, which could suppress the heterogeneous nucleation, broaden the growth window, and benefit larger growths of REBCO bulks.

**ACKNOWLEDGEMENTS**

The authors are grateful for financial support from the NSFC (Grant No. 51011140073 , 51172143 and 51072115), the MOST of China (Grant No. 2012CB821400) and the SSTCC (Grant No. 10JC1406800). We also appreciate Theva GmbH for offering excellent thin films.

[1] H. T. Ren, L. Xiao, Y. L. Jiao and M. H. Zheng, Physica C **412–414**, 597-601 (2004).

[2] Y. Shi, N. H. Babu and D. A. Cardwell, Supercond. Sci. Technol. **18** L13 (2005)

[3] X. Yao, K. Nomura, Y. Nakamura, T. Izumi and Y. Shiohara, Journal of Crystal Growth **234**, 611-615 (2002)

[4] C. Y. Tang, X. Yao, J. Hu, Q. L. Rao, Y. R. Li and B. W. Tao, Supercond. Sci. Technol. **18** L31 (2005)

[5] Y. Y. Chen, T. F. Fang, S. B. Yan and X. Yao, "Substrate effect on thermal stability of superconductor thin films", submitted

[6] M. Muralidhar, M. Tomita, K. Suzuki, M. Jirsa, Y. Fukumoto and A. Ishihara, Supercond. Sci. Technol. **23**, 045033 (2010)

[7] S. B. Yan, L. J. Sun, T. Y. Li, L. Cheng and X. Yao, Supercond. Sci. Technol. **24**, 075007 (2011)

[8] R. Semerad and J. Knauf, "High quality NdBCO films with thin YBCO buffer layers grown by reactive co-evaporation", Theva GmbH, 1-MA-P22

[9] C. Cai, K. Tachibana and H. Fujimoto, Supercond. Sci.Technol. **13**, 698 (2000)

[10] G. Li, M.H. Pu, X.H. Du, R.P. Sun, H. Zhang, Y. Zhao, Physica C **463–465**, 589–593 (2007)

[11] L. J. Sun, W. Li, Sh. F. Liu, T. Mertelj and X. Yao, Supercond. Sci. Technol. **22**, 125008 (2009)

[12] L. Cheng, C. Y. Tang, X. Q. Xu, L. J. Sun, W. Li, X. Yao, Y. Yoshida and H. Ikuta, J. Phys. D: Appl. Phys. **42** 175303 (2009)

[13] M. Muralidhar, K. Suzuki, A. Ishihara, M. Jirsa, Y. Fukumoto and M. Tomita, Supercond.

Sci. Technol. **23**, 124003 (2010)

[14] M.L. Griffith, R.T. Huffman and J.W. Halloran, J. Mater. Res. **9**, 1633, (1994)

Table 1. Various kinds of REBCO bulks seeded by the NdBCO/YBCO/MgO thin film and the $T_{max}$ of each process we succeeded in.

| RE123 systems | $T_{max}$ ( °C) | Mixed RE123 systems | $T_{max}$ ( °C) |
|---|---|---|---|
| SmBCO | 1115 | | |
| GdBCO | 1120 | Ag-GdBCO | 1115 |
| NdBCO | 1117 | Ag-NdBCO | 1115 |

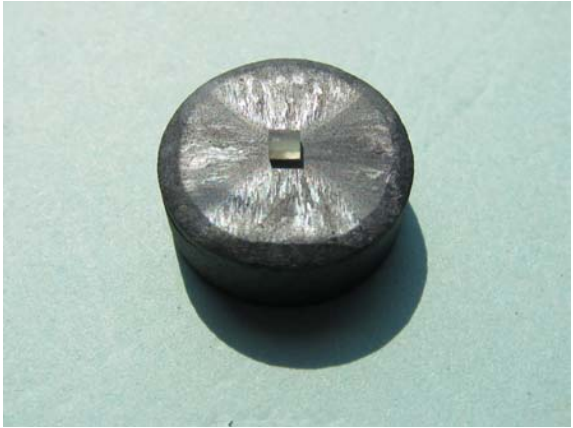

Figure 1. The as-grown NdBCO bulk seeded by NdBCO/YBCO/MgO thin film, using a Tmax of 1117 °C.

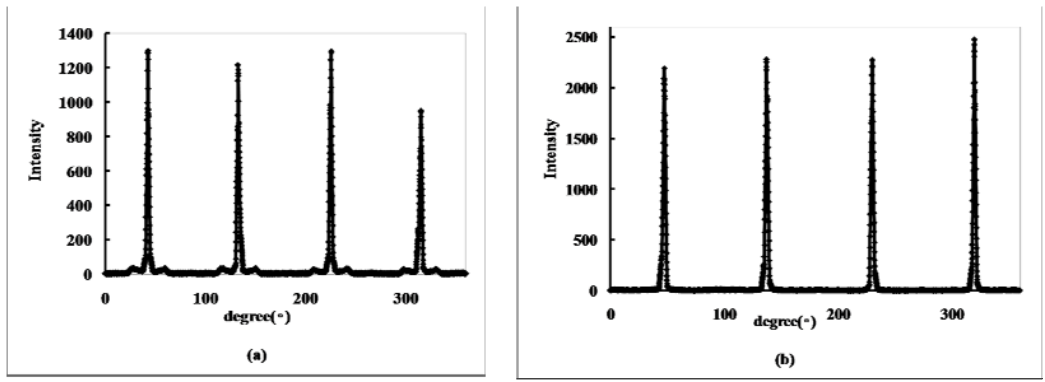

Figure 2. X-ray pole figure of: (a) NdBCO/MgO thin film; (b) NdBCO/YBCO/MgO thin film with a 25-nm YBCO buffer layer.

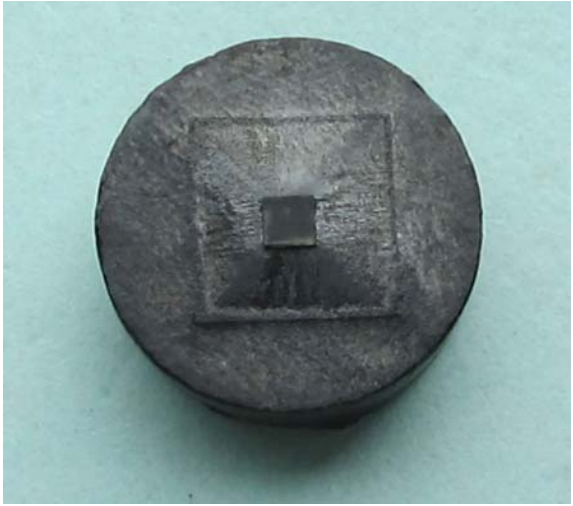

Figure 3. A GdBCO single grain seeded by the NdBCO/YBCO/MgO thin film using a $T_{max}$ of 1120 °C.

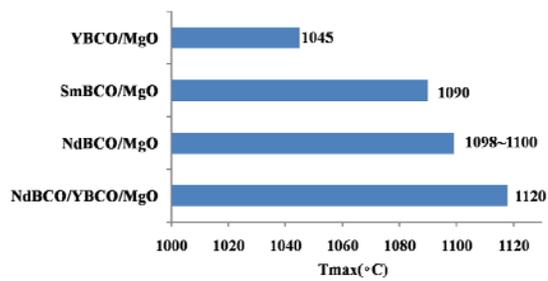

Figure 4. The types of film seeds that have been practically used and the $T_{max}$ they have been reported in the MT processes.